\documentclass{cta-author}
{}
{}
{}
\usepackage{etoolbox}
\makeatletter
\def\ps@pprintTitle{%
   \let\@oddhead\@empty
   \let\@evenhead\@empty
   \let\@oddfoot\@empty
   \let\@evenfoot\@oddfoot
}
\makeatother
\usepackage{placeins}
 \usepackage{url}
 \usepackage{comment}
\usepackage{listings}
\usepackage{setspace}
\usepackage{xcolor,pict2e}
\definecolor{eclipseStrings}{RGB}{42,0.0,255}
\definecolor{eclipseKeywords}{RGB}{127,0,85}
\colorlet{numb}{magenta!60!black}

\lstdefinelanguage{json}{
    basicstyle=\normalfont\ttfamily,
    commentstyle=\color{eclipseStrings}, 
    stringstyle=\color{eclipseKeywords}, 
    numbers=left,
    numberstyle=\scriptsize,
    stepnumber=1,
    numbersep=8pt,
    showstringspaces=false,
    breaklines=true,
    frame=lines,
    backgroundcolor=\color{white}, 
    string=[s]{"}{"},
    comment=[l]{:\ "},
    morecomment=[l]{:"},
    literate=
        *{0}{{{\color{numb}0}}}{1}
         {1}{{{\color{numb}1}}}{1}
         {2}{{{\color{numb}2}}}{1}
         {3}{{{\color{numb}3}}}{1}
         {4}{{{\color{numb}4}}}{1}
         {5}{{{\color{numb}5}}}{1}
         {6}{{{\color{numb}6}}}{1}
         {7}{{{\color{numb}7}}}{1}
         {8}{{{\color{numb}8}}}{1}
         {9}{{{\color{numb}9}}}{1}
}

\begin{document}

\title{SLA Conceptual Model for IoT Applications}

\author{\au{Awatif Alqahtani$^{1\corr}$}, \au{Ellis Solaiman$^{2}$}, \au{Rajiv Ranjan$^{2}$}}
\address{\add{1}{Computer Science and Engineering, College of Applied Studies and Community Service, King Saud University, Riyadh, SA}
\email{aqahtani1@ksu.edu.sa /}\\
\add{2}{Computing School, Newcastle University, Newcastle, UK}
\email{aqahtani1@ksu.edu.sa/}
{ellis.solaiman@ncl.ac.uk}}

\begin{abstract}
Since SLAs specify the contractual terms that are formally used between consumers and providers, there is a need to aggregate QoS requirements from the perspectives of Clouds, networks, and devices to deliver the promised IoT functionalities. Therefore, the main objective of this chapter is to provide a conceptual model of SLA for the IoT as well as rich vocabularies to describe the QoS and domain-specific configuration parameters of the IoT on an end-to-end basis. We first propose a conceptual model that identifies the main concepts that play a role in specifying end-to-end SLAs. Then, we identify some of the most common QoS metrics and configuration parameters related to each concept.
We evaluated the proposed conceptual model using a goal-oriented approach, and the participants in the study reported a high level of satisfaction regarding the proposed conceptual model and its ability to capture main concepts in a general way.

\end{abstract}
\maketitle

\vspace{2cm}
Note:\\  
This paper is a postprint of a paper submitted to and accepted for publication in 
[Managing Internet of Things Applications Across Edge and Cloud Datacenters. IET] and is subject to Institution of Engineering and Technology Copyright. The copy of record is 
available at the IET Digital Library

\vspace{1cm}
\maketitle

\onecolumn

\section{Introduction}\label{ch3-Intro}

IoT applications are mostly time-sensitive applications. Thus, it is important to consider when data needs to be collected, what the next processing step is, and where to process each step. Furthermore, the associated QoS requirements for each step should be specified in an unambiguous way. Therefore, there is a need to aggregate QoS requirements from the perspectives of Clouds, networks and IoT devices layers to deliver the promised IoT functionalities at the required quality level, as agreed upon within the SLA.\\

Consumers of IoT-based services have certain levels of expectations with respect to the quality of service provision. Therefore, it is important for an SLA to specify consumer requirements in relation to provider capabilities. The SLA provides a guarantee procedure that consumers' requirements will be met. However, IoT applications are mostly time sensitive. For example, in IoT-based emergency response (ER) applications, data from deployed sensors must be received and analysed immediately and accurately. Any delay in data transfer is unacceptable. Other examples are the prompt response required for natural disasters such as earthquakes, floods and tsunamis [278]. Therefore, for such applications and others, ensuring that consumer requirements are accurately and unambiguously specified within SLAs is crucial. Accurately specified SLAs are contracts that can form the basis of a strategy to regulate and automatize transactions and activities between interacting parties (service providers and consumers).\\

\begin{figure*}[thpb]
      \centering
\includegraphics[width=1\linewidth ]{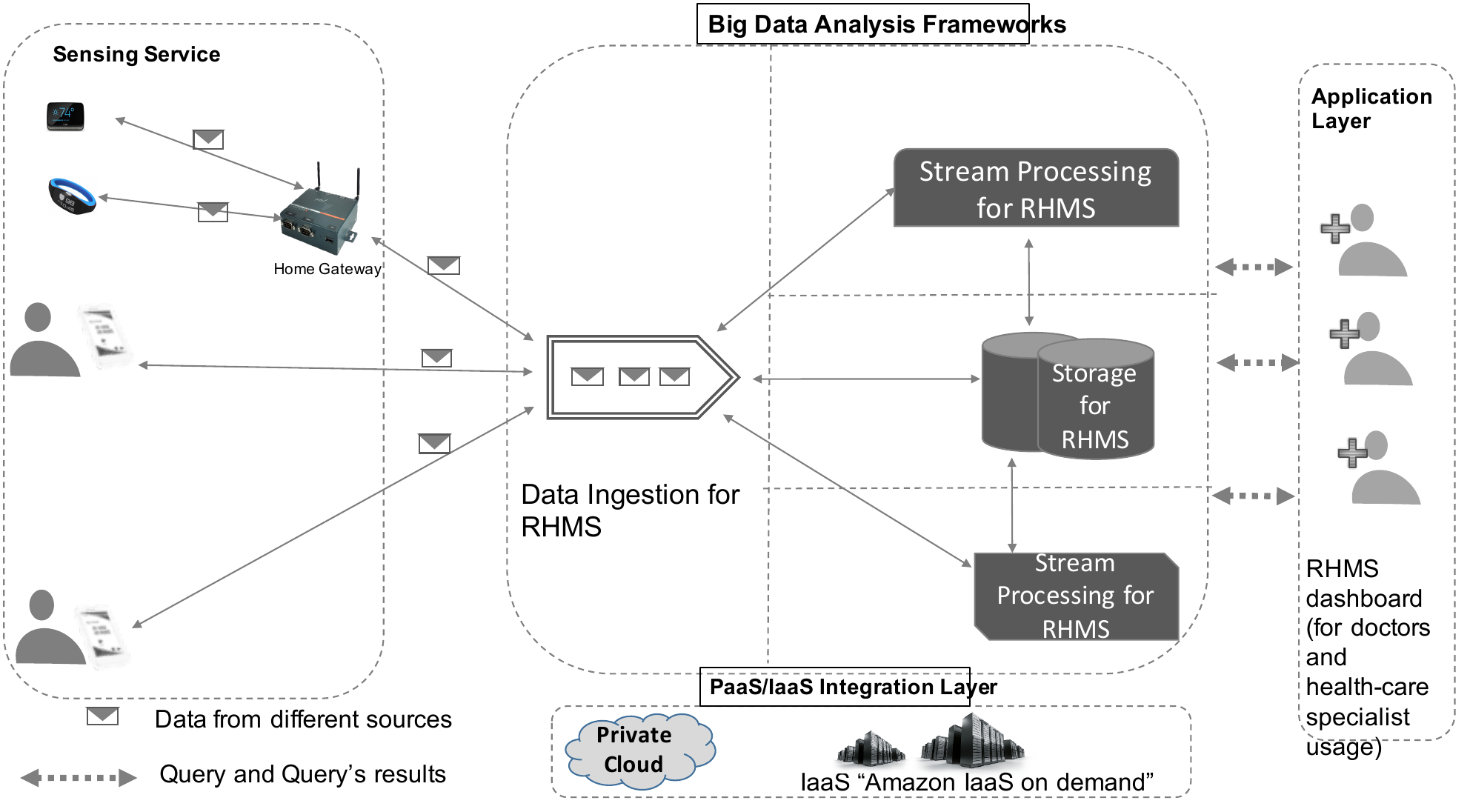}
      \caption{cooperated layers to deliver RHMS}
      \label{fig:RHMSEX}
   \end{figure*}

For further illustration, let us consider a remote health monitoring service (RHMS), in which patient data are collected from different sources (e.g., heartbeat sensors, smart cameras and mobile accelerometers) (Figure \ref{fig:RHMSEX} ). The filtered data are then transferred to a data-processing platform within the Edge/Cloud (depending on the required level of processing capabilities and/or storage capacity) layer for further analysis.
However, since the time factor is very important for this type of application, any unpredicted delay in one or more of the data flow stages (e.g., collecting, transferring, ingesting, analysing) will affect the accuracy and suitability of the actions taken. Therefore, the performance of an RHMS relies not only on the correctness of the provided functionalities but also on the quality of the offered services across the Edge and/or Cloud computing environments. Therefore, SLAs undoubtedly need to consider requirements across all layers of the Edge and/or Cloud environment, for example, at what rate data should be collected, transferred and ingested and how fast and accurate the analysis should be.\\

Traditional SLAs that focus on availability and reliability are not enough for IoT applications due to the need for strict SLA guarantees (of functions such as accuracy and speed of event of interest detection) \cite{chapterbook23}. Furthermore, having an individual SLA management mechanism for each layer of the IoT is inadequate because of the huge dependency across layers \cite{chapterbook24}. Thus, within an end-to-end SLA, there is a need to express constraints/policies that determine which data can be processed within the Edge data centres as well as which data need to be exported to be processed/analysed in Cloud data centres under certain constraints. Therefore, specifying the contractual terms of an SLA on an end-to-end basis is important not only to specify the end-to-end QoS requirements but also to assure consumers that their QoS requirements will be observed across computing environments to deliver services that match their expectations. This will aid service providers in operating their services at an adequate level, which will then increase consumers' trust, as it protects their rights if they encounter any damage during the contract period \cite{chapterbook31,chapterbook32}.\\

When specifying SLA terms on an end-to-end basis within a formal syntax language, standardizing the vocabularies used to describe the offered/requested services is crucial. With the multi-layered nature of IoT applications, there is a possibility of having more than one provider. Having multiple providers is a serious issue that leads to the need to standardize the terminologies used within the SLA to avoid ambiguity. For example, within the Cloud environment, there is a lack of standardized vocabularies in expressing SLAs. For example, availability is expressed differently by well-known Cloud providers: Amazon EC2 offers availability as a monthly uptime percentage of 99.95 per cent, Azure offers availability as a monthly connectivity uptime service level of 99.95 per cent, and GoGrid offers a server uptime of 100 per cent and an uptime of the internal network of 100 per cent \cite{sys206}. Furthermore, within the Edge environment, sampling rate \cite{chapterbook34} and sampling frequency \cite{chapterbook35} are used interchangeably to describe the rate at which a sensor sends data. Indeed, unifying metrics and terminologies as well as proposing a taxonomy will lead to a well-designed SLA, which in turn will provide a successful interaction between consumers and providers. Therefore, standardizing the vocabularies used to describe the offered services and the requested services may play a significant role in minimizing the ambiguity between cooperating parties. \\

A consumer who wishes to start an SLA must first select a service provider/s. Selecting service provider/s can be a challenging process, especially when considering the multi-layered nature of the IoT. To illustrate, consider the RHMS scenario, in which the IoT application administrators would aim to find the best set of providers that match their requirements. Since IoT applications have a multi-layered architecture, IoT administrators need to consider different categories of providers (e.g., network provider, Cloud provider) and find the best candidate for each category. Most popular Cloud providers (e.g., AWS, MS Azure, Oracle) currently provide descriptive take-it-or-leave-it SLAs for their services. When consumers need to compare such SLAs from different providers to select the most suitable, they must evaluate them manually \cite{ch219}. IoT applications can potentially be much more complex than Cloud applications, and such a comparison therefore becomes more difficult. Therefore, standardizing the vocabularies used to describe the QoS of the offered services and the requested services can be a first step towards enhancing the process of selecting service providers using certain search criteria. \\

 As a result, we attempt to contribute to the SLA of the IoT  by proposing a conceptual model that captures the knowledge base of IoT-specific SLAs.\\
 
This chapter contributes to the SLA for the IoT by:
\begin{itemize}
\item Proposing an SLA conceptual model
\item Introducing key concepts of SLA for IoT and the related vocabulary terms that can be used for specifying QoS and configuration parameters;
\item Evaluating the proposed conceptual model using a questionnaire-oriented approach from the domain experts' point of view.
\end{itemize} 
In the following text, Section \ref{conceptual} introduces the proposed conceptual model for IoT applications. Then, Section \ref{sec:KeyConcepts} presents the vocabulary terms that can be part of an SLA to reflect the QoS and configuration requirements. We evaluate the proposed conceptual model in Section \ref{sec:eval}.

\section{An End-to-End SLA Conceptual Model for IoT Applications} \label{conceptual}

An end-to-end IoT ecosystem includes  components through which application data flows. Components can include services (e.g., a sensing service or real-time analysis service), infrastructure resources (e.g., IoT devices, Edge resources, and Cloud resources) and/or humans. In end-to-end SLA, it is important to consider the requirements for all of the services and infrastructure resources involved to deliver the IoT application. Considering an SLA on an end-to-end basis is essential because establishing the SLOs of both services and infrastructure resources has an impact on establishing SLOs at the application level. For example, in an RHMS, an SLO (SLO$_{app1}$) for urgent case detection, which requires a response within less than \texttt{Y} time units, is an SLO at the application level, and it involves many activities, such as analysing real-time data. Analysing real-time data requires a stream-processing service at an acceptable level of latency, and if the stream-processing service exceeds this level, then SLO$_{app1}$ might be violated. \\

 As a result, we propose a conceptual model that captures the knowledge base of IoT-specific SLAs.  The conceptual model expresses the key entities of the IoT ecosystem and the relationships among those entities within the SLA context. Due to the lack of a standard IoT architecture, we refer to our reference IoT architecture as presented in \cite{awa} to identify the main concepts and the relationships among them.

\begin{figure*}[thpb]
      \centering
\includegraphics[width=.9\linewidth]{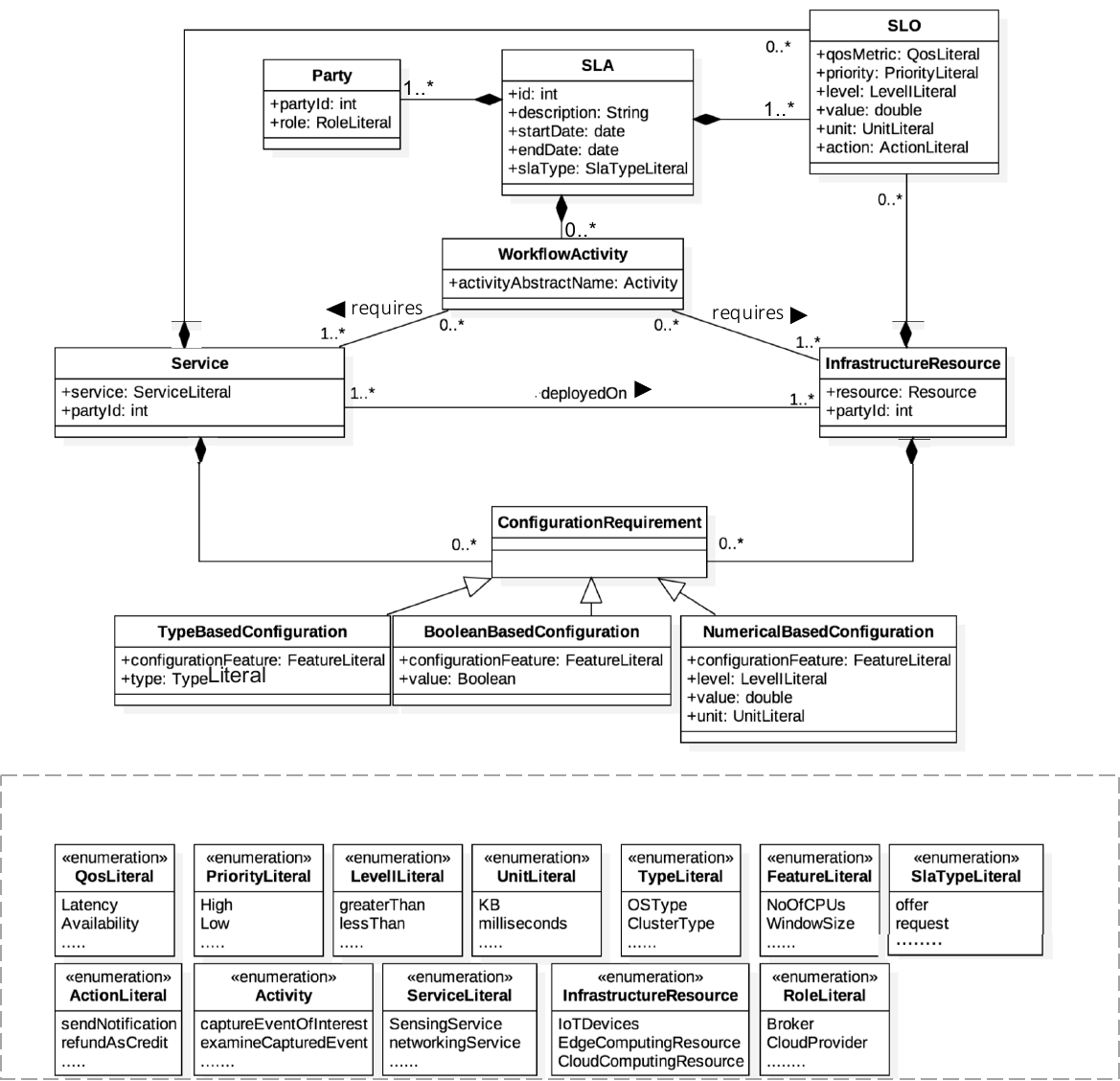}
      \caption{SLA conceptual model for IoT applications that captures the key entities of an SLA and the corresponding relationships \cite{awa}}
      \label{uml}
   \end{figure*}

Figure \ref{uml} presents our conceptual model. In the following section, we describe the concepts covered in the conceptual model and give a brief discussion of the relationships associated with these concepts.
\FloatBarrier
 
 \FloatBarrier

 The conceptual model is composed of the following entities:
\begin{enumerate}
\item SLA: 
The SLA includes basic data, such as the title of the SLA, the corresponding ID, the type of application (i.e., smart home, smart health, etc.) and the start and end dates.
\item Party: The part describes an individual or group involved in the SLA and usually includes a named company or a judicial entity \cite{8}. For example, in an RHMS, the parties could be the hospital management group, patient, network provider and Cloud resource providers. 
\item SLO: The SLO provides the quantitative means to define the level of a service that a consumer can expect from a provider.  It expresses the objective(s) of an agreement for both the application and any involved services and infrastructure resources. The SLO quantifies the required value of a QoS metric. For example, an SLO (at the application level) of an RHMS could be the response to urgent cases within \texttt{Y} time units. The QoS metric in this example is \texttt{response time}, and the constraint is less than \texttt{Y} time units. Furthermore, SLO parameters can be used to specify an SLO for low-level services; for example, for a data-ingestion service, an SLO can be: \emph{ingest data with latency less than \texttt{Z} time units}. For an infrastructure resource such as the CPU of a VM, an SLO can be: \emph{CPU utilization is greater than 80 per cent}.  
\item Workflow Activity:
IoT applications have certain activities that must be considered as part of the application requirements to function correctly. For example, in an RHMS, the possible workflow activities include capturing interesting data, analysing real-time data, and storing interesting results in a database (e.g., SQL or NoSQL).
In general, workflow activities mostly include:
\begin{itemize}

\item Capturing events of interest
\item Examining the captured events of interest on the fly
\item Filtering the captured events of interest
\item Aggregating the captured events of interest
\item Ingesting data from one or more data resources
\item Small-scale real-time data analysis 
\item Large-scale real-time data analysis 
\item Large-scale historical data analysis 
\item Storing structured data
\item Storing unstructured data
\end{itemize}
\item Service: This concept covers the main services that can be run/deploy to perform certain functionality. To achieve SLOs at the application level, it is important to establish adequate cooperation between particular services under the SLO constraints. For example, in an RHMS, to detect urgent cases within \texttt{Y} time units, it is necessary to transfer data from sensors to the ingestion service using networking service and to process data on the fly using stream-processing service. \\

Here, we list the most common services that can cooperate to deliver the SLOs of an IoT application.  
\begin{enumerate}
\item Sensing service: This service collects data using IoT devices and sends the collected data through a communication protocol to a higher layer. The sensing service specifies type of data and when to collect the data. For example, in an RHMS, a heartbeat sensor attached to the chest and an accelerometer as a hand-wrist device reflect a patient’s health state continuously or periodically based on what has been specified within the SLA for the service. 
\item Networking service: 
This service transfers the collected data from one layer to another. For example, in an RHMS, a home gateway uses the network to deliver collected data to the Cloud for further analysis under certain bandwidth requirements. 
\item Ingestion service: 
This service ingests data from many data producers and then forwards the data to subscribed/interested destinations such as storage and analysis services under certain requirements, such as throughput limits. 
\item Batch processing service: 
	A batch-processing service receives data from resources such as ingestion layers, appends them to a master data set and then provides batch jobs. For example, in an RHMS, to identify urgent cases, it is important to run machine-learning algorithms on historical patient records to recognize patterns regarding certain health issues and establish a predictive model. The predictive model can be used later with the real-time data of current patients to detect particular health issues. Batch views can be computed/queried within response time constraints, as specified by consumers/subscribers. 
\item Stream Processing Service:
This service processes incoming data from data resources such as an ingestion service to complete real-time tasks. For example, collected data are processed on the fly, and if the analysis shows an abnormality such as a high heart rate, then appropriate action is required, such as sending an ambulance. However, to exploit real-time data to the greatest extent possible, consumers/subscribers can specify certain requirements such as the maximum acceptable latency for computing/querying real-time views.
\item Machine learning service:
This is a service that applies different machine-learning algorithms for different purposes, such as providing predictions and extracting different dimensions of knowledge from collected data. For example, the service may apply a machine-learning algorithm on historical data collected from previous flood incidents as training data to create a model for predicting a flood based on incoming real-time data; this approach may prevent disasters from happening or at least reduce damage by warning people in advance. 
\item Database service (SQL and NoSQL databases): 
This service is used by ingestion, batch and stream-processing services to store or retrieve data, batch views and real-time views as intermediate or final data sets. Consumers can provide their requirements, such as setting a query response time, and specify whether data encryption is required. 
\end{enumerate}

\item Infrastructure resource: This concept covers the required hardware for computations, storage and networking, which is essential for deploying/running the abovementioned services. The infrastructure resources can be IoT devices, Edge resources and Cloud resources. \\

\begin{enumerate}
\item IoT devices: These devices include intelligent devices/objects with the ability to execute actions and reflect the physical world. 
\item Edge resources: These resources allow data processing to take place at the Edge of a network and include various types of resources, such as border routers, set-top boxes, bridges, base stations, wireless access points and Edge servers\cite{10Buyya}. These examples of Edge resources can be used to support Edge computations with specialized capabilities \cite{10Buyya}.
\item Cloud resources: 
These resources provide infrastructure as a service (IaaS) and are mostly located geographically far from the source (e.g., IoT device)\cite{10Buyya}.
\end{enumerate}
\end{enumerate}

The relationships among the above entities, which are depicted in the conceptual model (Figure \ref{uml}), are as follows. There is a one-to-many relationship between the SLO and the SLA entities to express the SLO constraints at the application level. Therefore, each SLA entity has a composite relationship with an SLO entity. An example of SLO at the application level could be the \emph{end-to-end response time of an application should be less than \texttt{Y} time units}. Furthermore, an SLA has a composite relationship with parties since parties are responsible for providing a service, using a service and/or playing third-party roles (e.g., to monitor a service). \\

Additionally, an IoT application has a set of workflow activities (e.g., capture an event of interest (EoI) or analyse real-time data) that cooperate to deliver the application. Therefore, there is a composite relationship between the \texttt{SLA} and \texttt{WorkflowActivity} entities. 
Each workflow activity requires a service (e.g., a sensing service, networking service, or stream processing service). Each service is deployed on one of the infrastructure resources (for example, an IoT device, an Edge resource, or a Cloud resource). Furthermore, each one of the services (e.g., sensing is a service) and infrastructure resources (e.g., VM is an infrastructure resource) can have an SLO/SLOs. For example, maximizing the level of data freshness could be an SLO for sensing services, and maximizing CPU utilization could be an SLO for a VM). Furthermore, each one of the services and infrastructure resources can have zero or more configuration parameters (e.g., the sample rate of the sensing service and number of CPUs per VM of an infrastructure resource). Therefore, there is an association relationship between \texttt{InfrastructurResource} and \texttt{Service} and composite relationship between \texttt{InfrastructurResource}, \texttt{Service}, \texttt{SLO} and \texttt{ConfigurationRequirement} entities. The dashed rectangle has a set of predefined data types which are defined as enumeration.\\

\section{Vocabulary Terms of the Configuration Parameters and QoS Metrics} \label{sec:KeyConcepts}
In this section, we cover the "service" and "infrastructure resource" concepts  of the considered services and infrastructure resources, with their sub-classes depicted in Figure \ref{uml}, in depth. We describe  the "service" and "infrastructure resource" concepts bellow with some of the related QoS metrics and configuration parameters.\\

 We search the literature to collect vocabulary terms that are related to the QoS metrics and configuration parameters. The reason behind considering the terms related to configuration parameters is the strong correlation between the QoS and configuration parameters; for example, the data publishing rate, as a configuration parameter, affects the data freshness as a QoS metric. This step comes after specifying the main components of the IoT reference architecture; then, the vocabulary terms that can be used to express consumer requirements are identified for each component. We believe that identifying domain-specific terms is the first step in providing unified/standardised vocabularies to mitigate the risk that can be caused by the ambiguity between the different providers who cooperate to deliver an IoT application  (Section \ref{sec:services} and Section \ref{sec:InfrastructureResources}). 
\subsection{Infrastructure Resources} \label{sec:InfrastructureResources}
Infrastructure resources include the name of the infrastructure resource used to deploy/host a service. An infrastructure resource includes the following components.

\subsubsection{IoT devices}
IoT devices consist of heterogeneous sets of devices such as sensors that capture information about the physical world by sensing some physical parameters of interest or detecting other smart objects \cite{37}. There are several QoS metrics related to perception layers, such as the optimum number of active sensors, sensor quality, energy consumption, data volume, trustworthiness, coverage, and mobility \cite{chapterbook10,chapterbook34,chapterbook35}. \\

However, some of these identified metrics may be inconsiderable for a single IoT device \cite{chapterbook35}, and these metrics are not trivial when considering the number of deployed devices that cooperate to deliver a service. For example, a sensor with a power consumption value equal to 0.9 watt-seconds seems fine, but when a network of hundreds of sensors is deployed, the cumulative value of the power consumption makes a difference \cite{chapterbook35}. \\

IoT communication protocols can be varied in their communication range, bandwidth and power consumption. Thus, it is important to consider support for different types of protocols, and the most appropriate type that satisfies the application requirements should be selected. For example, if the power consumption is the most important key requirement, then ZigBee, as a communication protocol that can be characterised as a low power consumption protocol \cite{zigbee}, should be used; alternatively, WiMax is a protocol that provides a high communication bandwidth. Therefore, it is essential to select devices that support the preferred communication protocol. Some of the available communication protocols are Bluetooth, WiFi, ZigBee, 6LowPAN, Cellular, ANT, Z-Wave, Thread, WiMax and NFC \footnote{See \cite{comparative} for further details and a comparison of communication protocols}. Table \ref{table:IoTDevice} lists a number of vocabulary terms that can be used to express the requirements related to IoT devices.

\FloatBarrier \begin{table}[h]
\centering 
\caption{Terminology/vocabulary definitions related to IoT devices}
\small

\begin{tabular}{p{4cm}p{11cm}}
\hline
\textbf{Terminology} & \textbf{Definition/Description /Example} \\
\hline
Device accuracy & Description of how well the device reflects an
interesting event. \\
Device precision & Description of how precisely the device reads an
interesting event in a stable manner. \\
Type of device	& For example, sensors and RFID tags.\\
Number of devices & The number of devices. \\
Mobility of devices &	Specification of whether the device is fixed or mobile (this feature affects network coverage).\\
Communication mechanism	&The mechanism of pushing/pulling data to/from the next layer. This mechanism can be a built-in hardware feature, a software feature or both.\\
Communication technology & The communication protocol with other devices that are supported, such as by WiFi and Bluetooth. This technology can be a built-in hardware feature, a software feature or both.	\\
Battery life & Battery life is a measure of battery performance and longevity, which can be quantified in several ways: as the run time on a full charge, as the milliampere hours estimated by a manufacturer, or as the number of charge cycles until the end of useful life. \\
Warranty period & The time period in which a purchased device may be returned or exchanged.\\
	Storage size &	The storage size of an IoT device that can be used to store data.\\

Memory capacity & The maximum or minimum amount of memory an IoT device has.\\
CPU capacity &  The capability and speed of a processor which reflects how many operations it can perform within a given amount  of time.  \\


\hline
\end{tabular}
\label{table:IoTDevice}
\end{table} \FloatBarrier

\subsubsection{Edge resources}
In an Edge layer, intelligent computation abilities are allocated to Edge resources (a gateway, server, etc.) to improve performance and reduce unnecessary data transfers to Cloud data centres. Edge resources contain sensitive personal and social data, and data management and control tasks are moved to the Edge to be managed in a secure and private manner \cite{cite6}. Edge resources mostly include border routers, set-top boxes, bridges, base stations, wireless access points, Edge servers, etc. These examples of Edge resources can be equipped to support Edge computations with specialised capabilities \cite{10Buyya}. A gateway typically links devices with the Cloud layer where data can be processed, stored and analysed. The IoT devices (e.g., sensors) can work without a gateway if the sensors have the ability to communicate directly with the Internet. In this case, sensors might have lightweight functionality  \cite{5lit}. Furthermore, for a more cost-effective approach with typical sensors that do not have gateway capability, it is possible to use many-to-one mapping, where many sensors can be connected to one gateway, which then increases the data transfer capability  \cite{5lit}.\\

Smart gateways can handle resource constraints on the processing power, power consumption and bandwidth of connected devices by allowing constrained devices to outsource some functionalities to the gateway. These gateways can be provided with local databases for temporarily storing sensed data, as well as enhancing data fusion, aggregation and internal device communication \cite{chapterbook28}. When specifying the QoS for an application, it is necessary to decide whether to deploy typical sensors and a gateway or a smart sensor. For example, using smart sensors (a smart sensor (with some processing capabilities) can behave as an IoT or an Edge resource) reduces the delay that is required for transferring data to the Cloud layer, which might be located at distance position, and the data can be processed within Edge resources instead of forwarding them to the next layer.\\

Some configuration parameters can affect the overall QoS of an IoT application. For example, the data publishing rate at the gateway is a concern because an increase in this rate might cause the broker network to be "overloaded", which then causes messages to be dropped \cite{ch3rPublishRate}. Another configuration parameter is the buffer size, which plays a significant role in the performance of an IoT gateway. For instance, \cite{ch3rBuffer} proposed a multi-threaded gateway and considered different values for different parameters, including different buffer sizes to enhance gateway performance when evaluating the proposed model. Table \ref{table:edgeRes} lists a number of vocabulary terms that can be used to express the requirements related to Edge resources. 
\FloatBarrier
\begin{table}[h]
\centering 
\caption{Terminology/Vocabulary definitions related to Edge infrastructure resources}
\small
\begin{tabular}{p{3cm}p{12cm}}
\hline
\textbf{Terminology} & \textbf{Definition/Description /Example} \\
\hline
Availability & The ratio of the time that the resource is functioning as expected and ready for use divided by the total run time.\\
Type of device & For example, a mobile, raspberry pi, or server devices.\\
Gateway throughput
& The amount of data transferred through the gateway
per second.\\
 Gateway delay & The delay in data collection from nodes. \\
 Publishing rate
& 	Specifies when data need to be sent.\\
Number of devices	& Total number of devices within the Edge infrastructure. \\
Mobility of devices	& Specification of whether a device is fixed or mobile (this feature affects network coverage).\\
Communication mechanism	&	The mechanism of pushing data to the next
layer or pulling data from the next layer; it can be a built-in hardware feature, a software feature or both.\\
Communication technology	&	 The communication protocols with other devices, such as the communication protocols based on WiFi and Bluetooth. Such protocols can be a built-in hardware feature, a software feature or both.	\\

Storage/buffer size & The buffer/storage size that can be used
to buffer/store data due to limited throughput for incoming data or to buffer/store data until delivery confirmation is received.\\

Memory capacity &	The maximum or minimum
amount of memory an Edge resource is capable of having.\\
CPU capacity &  The capability and speed of a processor which reflects how many operations it can perform within a given amount  of time.  \\

\hline
\end{tabular}
\label{table:edgeRes}
\end{table} 
\FloatBarrier
\subsubsection{Cloud resources}
Most Cloud data centres are distributed internally across several physical data centres. As a result, many Cloud providers not only provide fault tolerance for a single machine or single rack but also provide resilience for full data centre failures, which yields a high level of reliability. Cloud providers supply computer resources on an on-demand basis. This approach quickly enables (typically in minutes) an arbitrarily large number of computing nodes to be accessed with scale-up and scale-down possibility \cite{ch3Cloudonline}. \\

Cloud resources can have one or more than one SLO; for example, an SLO can be "CPU utilization should be more than 80 per cent". Furthermore, a Cloud resource can have a configuration parameter, such as a number of vCPUs. Most Cloud systems provide a variety of storage system functions, such as those for the storage bandwidth, size, cost, latency, and access control for different storage types, including local instance storage, distributed block storage, distributed file systems and object (blob) storage. These various services can lead to very different choices regarding software design depending on the system or application requirements \cite{ch3Cloudonline}. 
Table \ref{table:Cloud} lists a number of QoS and configuration parameters for Cloud resources. There are different types of instances, e.g., instances with more RAM versus more storage, or with specific hardware components, such as GPUs or FPGAs \cite{ch3Cloudonline}. 


\FloatBarrier \begin{table}[h]
\centering 
\caption{Terminology/Vocabulary definitions related to Cloud infrastructure resources}
 \small

\begin{tabular}{p{4cm}p{11cm}}
\hline
\textbf{Terminology} & \textbf{Definition/Description /Example} \\
\hline
Availability &	The ratio of the time that the resource is functioning\\
& as expected and ready for use divided by\\
& the total run time.\\ 
CPU utilization	 & Percentage representing how the CPU is being utilized.\\
Outage length	& The length that the resource is not available.\\
Throughput & The data transfer rate to and from a Cloud resource per second.\\
Storage size & Available disc space for data storage purposes.\\
Storage bandwidth &	Measure of the capacity to transfer data between a
service and storage.\\
Storage type &	Type of storage for a service (e.g., local SSD or local HDD).\\
Input/output storage operations & The specified number of input/output operations for storage. \\
Access protocols & Cloud access protocols. \\
Memory capacity & The memory capacity is the maximum or minimum amount of memory a computer or hardware device is capable of having or the amount of memory required for a program to run.\\
Network bandwidth & Network speed among the internal service nodes involved (e.g., 100BASE-T, 100BASE-SX).\\
vCPU capacity	&	The capacity of each virtual central processing unit (vCPU) which reflects how many operations a vCPU can perform within a given amount  of time. \\\\
No. of vCPUs	&	The number of vCPUs per VM. \\
No. of cores per VM	& The number of cores per VM.\\
Vertical scale-down limit &	The minimum number of CPUs if scaling is not automatic. \\
Vertical scale-up limit	& The maximum number of CPUs if scaling is not automatic \\

Horizontal scale-up limit	&	The maximum number of VMs if scaling is not automatic. \\
Horizontal scale-down limit	&	The minimum number of VMs if scaling is not automatic. \\ 

Replication factor & The number of copies of data that one wants
the cluster to maintain.\\
\hline
\end{tabular}
\label{table:Cloud}
\end{table} \FloatBarrier

\subsection{Service Concept} \label{sec:services}
To achieve SLOs at the application level, it is important to rely on adequate cooperation among some services under the SLO constraints. Therefore, we use the service concept to capture the name of the required services. A service has one or more SLO constraints and configuration requirements, including but not limited to those for sensing, networking, stream processing, batch processing, database management, and machine learning algorithm services. Each one of the previously mentioned services can have one SLO or more; for example, an SLO for stream processing services can be "minimizing latency to be less than 5 time units". Furthermore, each of the previously mentioned services can have configuration requirements; therefore, there is a relationship between the service and configuration requirement concepts. For example, a service such as stream processing can specify a requirement related to the “window size” (the window size is a configuration parameter). In the following section, we list the most common services that can cooperate to deliver an IoT application.

\subsubsection{Sensing Services} 
A sensing service is responsible for collecting data from IoT devices and sending the collected data through a communication protocol to another layer. The sensing service specifies the number of sensors, type of sensors, and sampling rate. In an RHMS, for example, to provide a sensing service, we need to specify the type of sensors associated with a patient, such as a heartbeat sensor attached to the chest and an accelerometer on the hand/wrist to reflect the patients' activities. A sensing service is associated with different parameters that play a significant role in the overall QoS of an IoT application.\\



For example, different applications require varying sampling rates depending on their criticality. The sampling rate determines the frequency at which an observed phenomenon is measured by a sensor (e.g., 5 Hz) \cite{chapterbook34}. Moreover, gaps in historical data can cause IoT applications to behave unexpectedly, which affects the final outcome and can lead to a bad user experience. Therefore, the IoT platform must attempt to maximize data freshness \cite{ch3r4}. The importance of the freshness parameter from the perspectives of both producers and consumers has been recently discussed in \cite{ch3r8}. The authors argued that for transient IoTs, both data of interest and data packets should have a certain freshness to perform accurate caching and retrieval operations. Additionally, old content is automatically discarded from data storage as a consequence of the freshness requirement\cite{ch3r7}. Moreover, data freshness is one of the security requirements in the IoT because if an attacker first captured data and resent them, the data will become old \cite{ch3r1}\cite{ch3r3}.\\

Another metric is data quality, which is a complicated metric since it relies on other metrics, such as data accuracy \cite{ch3r1part2}. Data accuracy, itself, is affected by data freshness and precision \cite{ch3r10}, reflecting the high dependence among metrics. Furthermore, application objectives such as reducing energy consumption and non-functional properties are interdependent. For example, increasing the sampling rate plays a significant role in enhancing data freshness, which in turn improves the information quality; however, this change decreases battery life (i.e., increases energy consumption).
Table \ref{table:sensing} lists some of the vocabulary terms that can be used to express the QoS constraints and configuration parameters relevant to sensing services.

\FloatBarrier \begin{table}[h]
\centering 
\caption{Terminology/Vocabulary definitions related to sensing services }
\small
\begin{tabular}{p{3cm}p{12cm}}
\hline
\textbf{Terminology} & \textbf{Definition/Description/Example} \\
\hline
Availability &	The ratio of the time that the service is functioning \\
& as expected  divided by the total run time.\\

 Data freshness
& The age of sensor data because data cannot always be transmitted in real time/near-real time\\
 Sampling rate & The rate at which a sensor measures an observed phenomenon (e.g., 5 Hz). Different applications require different sampling rates based on their criticality.\\
 Data accuracy & The error rate of data. It is possible to specify the average number of errors over a given time period.\\
 Data integrity & Data integrity reflects the degree to which data have been maintained or altered.\\
 Data type & e.g., Capturing weather temperature or  humidity.\\
 \\

\hline
\end{tabular}
\label{table:sensing}
\end{table} \FloatBarrier

\subsubsection{Networking Services}
Networking services are used for passing the collected data from one layer to another and provide a bidirectional connection for cases in which an instruction needs to be sent to one or more devices. For example, in an RHMS, gateways use the network to deliver collected data to the Cloud for further analysis. A networking service is also used when a command is sent back to a sensor, for example, to reconfigure the sampling rate, collect more data or check a patient's status. Thus, a network service is responsible for transferring data between an IoT and an Edge resource \cite{reffnote12}. Furthermore, in some cases, an IoT device has the ability to communicate without needing a gateway; in such a case, the networking service is used to immediately connect the device to Cloud services (e.g., ingestion service and/or stream processing service). The quality of the network plays a significant role in delivering the data within the acceptable time limit before data lose value. Therefore, considering the QoS requirements of the network layer is crucial. \\

QoSs have been extensively researched in the field of network communications and have well-defined and measurable characteristics, such as throughput, jitter or packet loss \cite{ch3r10}, which impact the network delay \cite{reff16} \cite{reff22} \cite{reff24} \cite{reff30} \cite{reff36}. 
Table \ref{table:networking} lists some of the vocabulary terms that can be used to express QoS constraints and configuration parameters that are relevant to networking services. 

\FloatBarrier
\begin{table}[h]
\centering 
\caption{Terminology/Vocabulary definitions related to networking services}
\small

\begin{tabular}{p{3cm}p{12cm}}

\hline
\textbf{Terminology} & \textbf{Definition/description/Example} \\
\hline
Availability &	The ratio of the time that the a network is fully operational as expected and ready for use  divided by the  period of time.\\ 
Link bandwidth 
& The maximum amount of data that can be transferred through a link
per second.\\
 
Network delay & The delay in data transmission.\\
Data-in rate & The amount of incoming data per time unit. \\
Data-out rate & The amount of outgoing data per time unit.\\
 	
Jitter
& The time delay variance between data packets over a network in milliseconds (ms).\\
Packet loss rate
& 	The ratio of the number of packets lost to
the total number of packets sent. Each packet has a deadline for execution, and if meeting this deadline is not possible, the scheduler tries to minimize the number of packets lost due to deadline issues.\\

 Data integrity & Data integrity reflects the degree to which data have been maintained or altered.\\
 
 \\

\hline
\end{tabular}
\FloatBarrier

\label{table:networking}
\end{table} \FloatBarrier

\subsubsection{Ingestion Services}
An ingestion service describes how data can be ingested from many data producers \cite{5} and then forwarded to subscribed/interested destinations, such as a storage service, analysis service and/or application. 
An ingestion service can be associated with different parameters, such as configuration requirements (e.g., the number of servers/nodes and compression/decompression support) and some SLO constraints (e.g., maximizing throughput and minimizing latency). \\

In ingestion service, data often come from a variety of sources, including web logs, databases, various kinds of applications, etc., making it hard to understand what sort of data the system will ingest. One alternative is to use big data (BD) software, which can collect and aggregate data from various sources. Projects such as Flume \footnote{http://flume.apache.org/} and Scribe \footnote{https://github.com/facebookarchive/scribe/wiki} enable the collection, aggregation and transfer of large quantities of log information from many distinct sources to a centralized data storage centre \cite{ch3rIngestion}.\\

Data retention is one of the parameters that service consumers need to specify to indicate how long data can be stored before they are deleted. Therefore, the data rate and data retention time are interdependent since they represent key factors related to resource storage. For example, in Kafka \footnote{https://kafka.apache.org/}, the data rate of a partition is the rate at which it generates information; in other words, it is the average size of the message multiplied by the amount of messages per second. The data rate indicates how much retention space is needed in bytes for a given amount of time to ensure retention. If there is a lack in knowledge regarding the data rate, the retention space needed to meet a time-based retention goal cannot be calculated properly \cite{ch3rIngestionKa}. \\

 Messaging systems provide some replication-related functionality to improve various factors, including reliability, fault tolerance, and accessibility for replicating data/messages on different servers. For example, replication is used by default in Kafka; even unreplicated topics are implemented as replicated topics \cite{ApacheKa}. Data encryption, data compression, and a delivery guarantee mechanism are application dependent, so if providing a low-latency solution is important, then the data encryption delivery guarantee mechanism will cause delays. Furthermore, if reliability is important, then providing a delivery guarantee mechanism that ensures that messages/data/requests are delivered using the ingestion service is crucial. In other cases, when throughput is highly prioritized over latency, data compression is a key concern. The available messaging systems provide compression, encryption and delivery guarantee mechanisms; as an example, Amazon Kinesis Data Firehose \footnote{https://docs.aws.amazon.com/firehose/latest/dev/what-is-this-service.html} enables the compression of information before it is delivered, and it supports the GZIP, ZIP, and SNAPPY compression formats \cite{AmazonK}. Amazon Kinesis Data Firehose, also, allows for data encryption using the AWS Key Management Service \cite{AmazonK}. RabbitMQ \footnote{https://www.rabbitmq.com/} and Kafka both offer long-lasting messaging guarantees. Both offer at-most-once and at-least-once guarantees, but in very restricted situations, Kafka provides precisely once guarantees \cite{RabbitMQ}.
 Table \ref{table:ingestion} lists some of the vocabulary terms that can be used to express some of the QoS constraints and configuration parameters that are relevant to ingestion services. 
\FloatBarrier
\begin{table}[h]
\centering 
\caption{Terminology/vocabulary definitions related to ingestion services}
\small

\begin{tabular}{p{4cm}p{11cm}}
\hline
\textbf{Terminology} & \textbf{Definition/Description /Example} \\
\hline
Availability &	The ratio of the time that the ingestion service is functioning as expected  \\
& divided by the  period of time.\\ 
	Throughput
& 	The amount of data transferred through the messaging platform per second.\\
 	Latency & 	The time required to process a single input/output transaction before forwarding it to its destination within the ingestion service framework. \\
 
	Data-in rate & The amount of incoming data per time unit.\\

 Data-out rate & The amount of data output per time unit.\\
 	
Data retention time limit & The limit of how long data can be
saved in the ingestion layer.\\
Publishing rate
& Rate at which data is sent to a message broker.\\
Storage size & 	The amount of storage that can be used to store data due to limited throughput constraints considering the amount of incoming data, to store data until delivery confirmation, or to store data during the specified retention time. \\
Replication factor & How many replicas can be stored.\\
Data compression  support & 	A Boolean value that expresses whether 
 	data can be compressed/ decompressed depending on the requirements.\\

 Data encryption support & 	A Boolean value that expresses whether data
can be encrypted/decrypted depending on the requirements.\\
 	Delivery guarantee mechanism
& 	It reflects if data have been delivered to the destination. It affects the workload if the type of delivery
guarantee mechanism requires sending an acknowledgement back to the data producer. \\
 Data integrity & Data integrity reflects the degree to which data have been maintained or altered.\\
Name of ingestion framework & e.g., RabbitMQ, Amazon Kinesis Data Firehose, Flume, Scribe \\
 
 \\

\hline
\end{tabular}
\label{table:ingestion}
\end{table} \FloatBarrier
\FloatBarrier
\subsubsection{Stream processing services}
A stream processing service refers to processing incoming data from different data sources and/or ingestion services to compute real-time views. Furthermore, real-time views can be combined with saved computed batch views using a database framework (such as Cassandra \footnote{http://cassandra.apache.org/}) to answer some questions that rely on both real-time views and batch views. In an RHMS, data can be collected using different sensors, such as wearable accelerometers, that can be augmented by distributed-motion sensors for activity recognition purposes \cite{19}. If the collected data show abnormality for a given activity, such as an elderly person falling down, then an appropriate action, such as sending an ambulance, is required. However, applications such as RHMSs rely on real-time data; therefore, any delay in data processing could cause the data to lose their value. \\

High throughput and low latency are very important QoS requirements in stream processing. If incoming data are not analysed in real or near-real time, then the action taken may not be appropriate since actions are based on data that are no longer considered real-time/near-real-time data due to the delay. Another important metric is data completeness, which "measures the percentage of incoming stream data that are used to compute the query results." \cite{ch3rStream}. To illustrate the concept of data completeness, consider a data stream with a number of incoming tuples. In the ideal case, the query should be performed using a large sliding window, e.g., containing 30 tuples; however, due to resource constraints, 15 tuples are sampled and used to execute the query, which represent 50 per cent of the 30-tuple window size. The sampling method decreases the query data completeness to 50 per cent \cite{ch3rStream}. Furthermore, another QoS metric is the miss ratio, which "evaluates the number of queries that are not completed within the given time constraints" \cite{ch3rStream}. \\

In addition, the single-point resource estimation is insufficient to handle stream processing workloads in which information flows endlessly through the operator graph and yields changes in performance and resource demands. Therefore, to illustrate the effects of certain configuration parameters on performance and resource usage, consider the work in  \cite{ch3rRajivStream} as an example. Khoshkbarforoushha et al  \cite{ch3rRajivStream} presented a novel method using mixed density networks, a mixed structure of neural networks and mixed models to estimate the resource usage of data stream processing workloads in the Cloud. To train the proposed model, a set of features was used as the model input; the set included the size of windows that can be expressed in time units (second) or tuple units (number), the sliding value of the window type, the average arrival rate of tuples (tuple/second) to query, the total number of nested sub-queries and the operator type. The set of features was customized based on the prediction goal because the impact varied with respect to the CPU and memory. A feature that is correlated with memory consumption may not be correlated with CPU usage. For example, the selection results for features suggest that the size of the window has an insignificant effect on the prediction of CPU use but a notable influence on the prediction of memory use \cite{ch3rRajivStream}.\\

Furthermore, the QoS requirements of stream processing are affected by other configuration parameters, such as the window size and query size; in addition, the choice of a stream processing framework affects the QoS. For instance, selecting a framework (such as Spark streaming) \footnote{https://spark.apache.org/streaming/} that stores data before processing affects the latency level; Apache storm \footnote{https://storm.apache.org/} can process data immediately with no need to store them first \cite{cite4}. Table \ref{table:streamprocessing} lists key terms/definitions related to stream processing services to express requirements for both QoS metrics and configuration parameters.\\

\FloatBarrier \begin{table}[h]
\centering 
\caption{Terminology/vocabulary definitions related to stream processing services}
\small

\begin{tabular}{p{4cm}p{11cm}}
\hline
\textbf{Terminology} & \textbf{Definition/Description /Example} \\
\hline
Throughput & 	The stream size processed per second.\\
Latency & The time required to process a single input/output transaction for a stream processing service. \\
Data completeness & Measurement of "the percentage of incoming stream data that are used to compute the query results." \cite{ch3rStream}\\
Miss ratio & "Miss ratios measure the percentage of queries that are not finished within the given deadlines." \cite{ch3rStream}\\
Time-based window size & 	The size of the window with respect
to the time required to process data that occur within the window. The window size can be over time or based on a number of records/messages.\\
Event-based window size & The size of the window based on a
number of events/records/messages within a given window.\\
Sliding window & Determines the length of the window and the portion of the range that is retrieved when the window moves forward; the intervals can overlap. This value can be time based, count based or based on a hybrid scheme.\\
Tumbling window & A series of fixed-sized, non-overlapping and contiguous time intervals.\\
Micro batch size & 	Specification of the size of data that need to be buffered first before being processed; however, in stream processing, data are not required to be stored first. It is better if data are processed in active mode, which means that data are processed as they arrive and not when they are pulled.\\
Data arrival rate & Specification of how many data points are expected to be received per second.\\
Write capacity & 	Specification of the capacity of writing in one go.\\
Read capacity & 	Specification of the capacity of reading in one go.\\
Replication factor & 	Expression of how many replicates can be stored.\\
Total number of queries & 	Specification of how many queries should be
considered.\\
Data Compression support & 	A Boolean value that expresses whether data    can be compressed/ decompressed depending on the requirements.\\
Data Encryption Support & 	A Boolean value that expresses whether data
can be encrypted/decrypted depending on the requirement.\\
 Data Integrity & Data integrity reflects the degree to which data have been maintained or altered.\\
 Name of stream processing & e.g., Spark streaming, Apache storm\\
 framework&\\

\hline
\end{tabular}
\label{table:streamprocessing}
\end{table} \FloatBarrier

\subsubsection{Batch Processing Services}
A batch processing service refers to receiving data from ingestion layers and/or other data sources, appending the data to the master dataset and then obtaining batch views; moreover, the computed batch views can be stored for inquiry purposes. Batch processing can be based on incremental algorithms or recomputation algorithms \cite{17} considering the type of job that needs to be accomplished. For example, in an RHMS, if hospital management is interested in recording some statistics regarding the detected urgent cases, one interesting statistic might be the total number of urgent cases that have been detected. The count function can then be applied using an incremental algorithm or recomputation algorithm. However, since new detected cases can quickly reach the previous total number of detected cases, an incremental algorithm could be more suitable. The reason for choosing an incremental algorithm in this case is that the total number can be calculated without considering the entire dataset; this process avoids the need for additional computational resources since it only requires an increment establishment step. \\

However, if the query must consider the whole dataset, such as for a query regarding the average age of people who have a certain health issue, then whenever new cases arrive, there is a need to recompute the average considering all of the recorded ages, which requires a recomputation algorithm. Selecting the appropriate algorithm is important. Recomputation algorithms require computational efforts/resources to handle the master dataset, while less computational resources are required for incremental algorithms. However, a recomputation algorithm is more robust since it is human-fault tolerant because batch views are continuously recomputed \cite{17}. \\

In batch processing services, the throughput and query response time are key QoS requirements in which users are interested.  The related terminology/vocabulary definitions are used to express configuration requirements (such as the number of map and reduce tasks and the batch size). Furthermore, the choice of which batch processing framework to select affects the QoS. For instance,  Hadoop \footnote{https://hadoop.apache.org/} is a powerful batch processing framework; however, it is not the appropriate choice when there is a need to apply machine learning algorithms because it requires data to be reloaded from the disc, which increases the latency; therefore, Apache Spark could be the ideal choice \cite{cite4}.\\

Furthermore, \cite{ch3Optimal} presented a mathematical model for the optimum number of map tasks in MapReduce resource provisioning to estimate the optimum number of mappers based on the resource specifications and data set size. The MapReduce library divides input data into several input splits. A map task reads an input split and processes the input split using the user-defined map function. The map function takes input key/value pairs and creates a set of pairs for an intermediate key/value. The mapper memory buffers the intermediate key/value pairs. If the size of the data set reaches the memory buffer threshold, intermediate key/value pairs are stored on the local disc and partitioned to reduce the task requirements using the hash function. The reduce tasks involve reading and sorting steps for the intermediate data and group data with the same key. Then, the key and intermediate value sets are sent as inputs to the reducer to be written to the reducer's memory, and the reduction function is invoked \cite{ch3verma2016big}. The output of the reduction function is concatenated and then written to the output file \cite{ch3Optimal}. The MapReduce model and Hadoop Open Source Implementation have proven effective for large data processing tasks and were inherently built for batch and processing jobs with high throughput requirements \cite{ch3Beyond}. Throughput, as a QoS metric, indicates the number of MapReduce jobs completed per time unit (e.g., minutes) \cite{ch3Throghput}. Furthermore, it should be noted that the number of map tasks can be used as a cost estimator, as applied in \cite{ch3Throghput}. 
Table \ref{table:batchProcessing} lists the terminology/vocabulary definitions related to expressing the QoS metrics and configuration parameters of batch processing services.
\FloatBarrier \begin{table}[h]
\centering 
\caption{Terminology/Vocabulary definitions related to batch processing services}
\small

\begin{tabular}{p{4.2cm}p{11cm}}
\hline
\textbf{Terminology} & \textbf{Definition/Description /Example} \\
\hline
Throughput & 	The number of batches that can be processed per second.\\
Response time & The time required to process a submitted job and receive a response. \\
Batch size & The limit on the size of each batch that is submitted to be processed.	\\
No. of batch jobs & The number of submitted batch jobs. \\
Process running frequency & Specification of how frequently the process needs to be run, e.g., twice per hour.\\
Max. memory of the map task & Amount of memory assigned to the map task.\\
Max. memory of the reduce task & Amount of memory assigned to the reduce task. \\
No. of mappers & The number of mappers.\\
No. of reducers & The number of reducers.\\

Write capacity & 	The capacity of writing in one step.\\
Read capacity & 	The capacity of reading in one step.\\
Replication factor & 	Expression of how many replicas can be stored.\\
Total number of queries & Expression of how many queries should be
considered.\\
Data compression support & 	A Boolean value that expresses whether data can be compressed/ decompressed depending on the requirements.\\
Data encryption support & 	A Boolean value that expresses whether data
can be encrypted/decrypted depending on the requirements.\\

 Data integrity & Data integrity reflects the degree to which data have been maintained or altered.\\
 Name of batch processing& e.g., Hadoop \\
 framework &\\

\hline
\end{tabular}
\label{table:batchProcessing}
\end{table} \FloatBarrier

\subsubsection{Machine Learning Services}
A machine learning service refers to a service that permits the use of various machine learning algorithms to predict the purposes and different dimensions of knowledge from the information collected. For instance, a machine algorithm can be applied to historical data collected from patients with heart attack incidents to obtain training data. Then, the training data can be used to create a model to predict heart attack cases based on incoming real-time data, which can prevent emergencies from happening or at least reduce patient damage by warning patients in advance. \\

In terms of practical needs, there are different QoS metrics, including speed, accuracy, price, etc., as in most topic detection and tracking (TDT) applications. Furthermore, different types of algorithms for machine learning affect accuracy and speed differently. The algorithm class reflects the type of algorithm, including classification, clustering, etc., whereas the algorithm name refers to the specific algorithm used, such as K-means, linear discriminant analysis (LDA) and naive Bayes \footnote{Refer to \cite{das2017survey} for further details about machine learning algorithms}. Different algorithms, even if they are from the same class, can have different impacts on the performance of a system. For example, some clustering algorithms, such as the K-means and Canopy algorithms, differ substantially in the speed of execution; specifically, K-means has more than one iteration, while Canopy has only one iteration \cite{ch3ML}. 
Table \ref{table:machinelearning} shows a list of the main QoS metrics and configuration parameters that are related to machine learning services. 
\FloatBarrier \begin{table}[h]
\centering 
\caption{Terminology/vocabulary definitions related to machine learning algorithm services} 
\small

\begin{tabular}{p{3cm}p{12cm}}
\hline
\textbf{Terminology} & \textbf{Definition/Description /Example} \\
\hline
	Accuracy		 & The accuracy of the analysis. \\
	Class of ML & The name of the class in which an algorithm is classified. For example, the supervised learning involves classification and regression algorithms, and the unsupervised learning class includes clustering and association algorithms.\\
	Name of ML algorithm & Specifies the name of the algorithm required, such as logistic regression, decision forest,
decision jungle, neural network, support vector machine, principal component analysis (PCA)-based anomaly detection, K-means, or naive Bayes. \\
	Way to run the ML algorithm	& Examples of this process are Sequential and MapReduce.\\

 Data integrity & Data integrity reflects the degree to which data have been maintained or altered. \\
\hline
\end{tabular}
\label{table:machinelearning}
\end{table} \FloatBarrier

\subsubsection{Database Services}
A database service can be used for data retrieval with different services, such as ingestion, batch and streaming services. The database service stores incoming data as an intermediate or final dataset, a set of computed batch views or set of computed real-time views. For instance, the incoming data can be initially stored, such as with HDFS in Hadoop, before any further processing. Then, the data can be retrieved for analysis or can be processed on the fly, and the derived results are stored in a database such as Cassandra.\\

There are different types of databases that are selected based on the purpose of the application and the required QoS. For example, in stream processing, data can be stored in databases that support low-latency read and write operations, whereas cases that require immutable data can use durable object storage platforms such as Amazon S3\footnote{https://aws.amazon.com/s3/}, which is preferable to other methods. Furthermore, to handle large amounts of data, a distributed storage platform is employed, such as the available open-source distributed database Druid\footnote{http://druid.io/}, which supports data ingestion as well as queries with low latency, and Apache HBase\footnote{http://hbase.apache.org/}, which supports the random and real-time reading/writing of large volumes of data. However, the selection of the appropriate platform to use is affected by some factors, such as the query response time \cite{cite4}.
Table \ref{table:database} lists some of the most common QoS metrics and configuration parameters of database services.
\FloatBarrier \begin{table}[h]
\centering 
\caption{Terminology/vocabulary definitions related to database services}
\small

\begin{tabular}{p{3cm}p{12cm}}
\hline
\textbf{Terminology} & \textbf{Definition/Description /Example} \\
\hline
Throughput & 	The queries that can be processed per second.\\
Response time & 	The time from when a user sends a request to when they receive a response. \\
Type of database & 	For example, SQL or NoSQL.\\
Type of NoSQL & For example, a key-value, document-based, graph-based, or
column-based NoSQL.\\
Read error rate & The number of errors associated with reading attempts per time unit (seconds).	\\
Cache hit ratio	 & The ratio of cache hits to misses, expressed as a percentage. A cache hit is when the data requested for processing are found in the cache memory. A cache miss is when the data requested for processing are not found in the cache memory. \\
Write error rate & Rate of errors associated with writing attempts per time
unit (seconds). \\
Write capacity & 	The capacity of writing in one step. \\
Read capacity & The capacity of reading in one step\\
Replication factor & 	Expression of how many replicas can be stored.\\
Compression support & A Boolean value that expresses whether data can be compressed/decompressed depending on the requirements. \\
Data encryption support & A Boolean value that expresses whether data can be 
encrypted/decrypted depending on the requirements.\\

 Data Integrity & Data integrity reflects the degree to which data have been maintained or altered.\\
 
 \\

\hline
\end{tabular}
\label{table:database}
\end{table} \FloatBarrier

\section{Evaluation} \label{sec:eval}
In this section, we present our evaluation approach to assess the proposed conceptual model. We have applied a goal-oriented questionnaire approach, and further details of the evaluation procedure and results are presented in the following sections.
\subsection{Experiment}
The main purpose of the conducted experiment is to evaluate the proposed conceptual model and to determine if it meets the relevant predefined goals: generality, based on the coverage of general concepts that are common in IoT applications; coverability, or the extent to which IoT application requirements are covered considering the main concepts that can be used within an SLA to express QoS constraints and configuration requirements; and accuracy, which reflects the extent to which the conceptual model is accurate and the overall satisfaction level.

\subsection{Participants}
The potential users of our proposed work are IoT administrators. Therefore, we performed an experiment in which the research interests/topics of participants were mainly related to IoT. The study was conducted with 15 participants; most of them were Ph.D. students who were working on topics related to IoT, such as remote health and smart city applications. Their research interests include Cloud computing, Edge computing and networking.

\subsection{Procedure}
The experiment was carried out following a well-defined procedure. First, one-on-one discussions were conducted in which each participant received an introduction to the SLA and the reference architecture of the IoT, and a presentation was given on the conceptual model. The participants were allowed to discuss and comment on the conceptual model. A use case was employed for scenario clarification purposes (an RHMS). At the end of this period, the participant was asked to submit a written version of the questionnaire in which there were four questions related to the conceptual model. Furthermore, there was a comment textbox to allow the participants to comment and make suggestions, provide criticisms, or give other feedback. The three questions related to the conceptual model were as follows:

\begin{itemize}
\item Overall, how satisfied or dissatisfied are you with our conceptual model?

\item To what extent does the conceptual model cover your requirements?


\item How satisfied are you with the conceptual model's generality?

\end{itemize}
\subsection{Experimental results}
\begin{figure*}[!htbp]
      \centering
\includegraphics[width=.9\linewidth ]{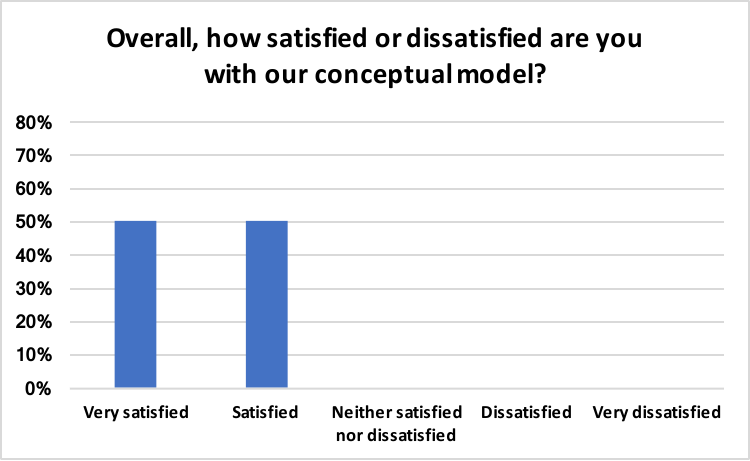}
      \caption{Results of the evaluation: Satisfaction}
      \label{fig:result1}
   \end{figure*}

\begin{figure*}[!htbp]
      \centering
\includegraphics[width=.9\linewidth ]{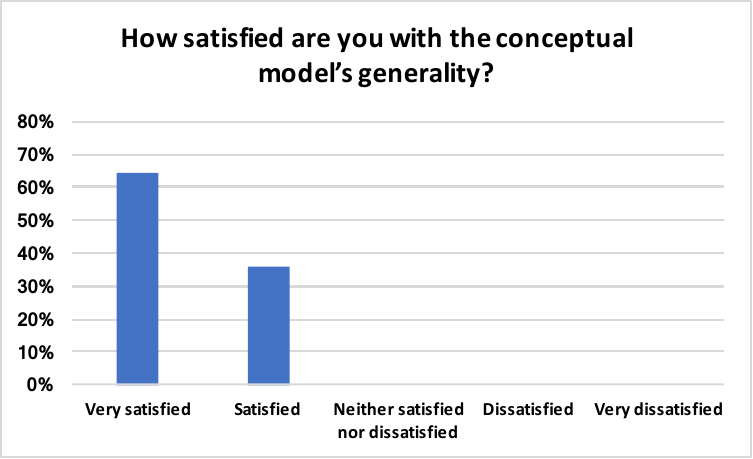}
      \caption{Results of the evaluation: Generality}
      \label{fig:result2}
   \end{figure*}

\begin{figure*}[!htbp]
      \centering
\includegraphics[width=.9\linewidth ]{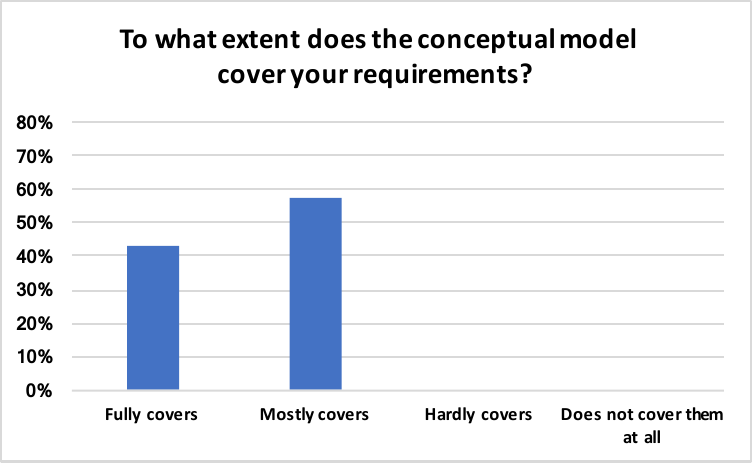}
      \caption{Results of the evaluation: Coverability}
      \label{fig:result3}
   \end{figure*}

Figures \ref{fig:result1}, \ref{fig:result2} and \ref{fig:result3} show the results based on participants’ answers in regard to the proposed conceptual model. Fifty percent of participants described their overall satisfaction level as satisfactory, while the other fifty percent were very satisfied. Regarding the coverability (capturing the main-related concepts) of the conceptual model, more than 40 per cent of participants answered that the model provided full coverage, and the rest of the participants answered “mostly covered”. Regarding the generality of the conceptual model, more than 60 per cent of participants were very satisfied, and the rest were satisfied. There were a few comments regarding concept names that describe resources, and it was suggested to change “resources” to "infrastructure resources". Furthermore, there was a comment regarding the presentation of the conceptual model as follows: "it would be better if it (the conceptual model) was represented in hierarchical view".

\FloatBarrier
\FloatBarrier
\FloatBarrier

   \FloatBarrier
\FloatBarrier

 \FloatBarrier

\section{Conclusion and Future work}
In this chapter, we tried to overcome one of the end-to-end SLA specification challenges related to the heterogeneity of key QoS metrics across the computing environment. We proposed a conceptual model for IoT-specific SLA. Then,  we identified domain-specific vocabulary terms that can be used as a starting point for an SLA specification considering both the QoS constraints and configuration parameters across layers. However, there is a limitation in the presented work related to the sample size of participants that evaluated the SLA conceptual model. However, the reason for the small size is that we sought participants with domain-specific knowledge, especially that we, mainly, were looking to review our conceptual model with experts.  In future work, we will try to extend the identified services and infrastructure resources and identify a list of vocabulary terms related to QoS metrics and possible configuration parameters. Furthermore, we will try to evaluate the proposed model with a larger sample size.

\bibliographystyle{unsrt}

\bibliography{references}

\begin{thebibliography}{10}

\bibitem{chapterbook23}
Meisong Wang, Rajiv Ranjan, Prem~Prakash Jayaraman, Peter Strazdins, Pete
  Burnap, Omer Rana, and Dimitrios Georgakopulos.
\newblock A case for understanding end-to-end performance of topic detection
  and tracking based big data applications in the cloud.
\newblock In {\em International Internet of Things Summit}, pages 315--325.
  Springer, 2015.

\bibitem{chapterbook24}
Awatif Alqahtani, Ellis Solaiman, Rajkumar Buyya, and P~Ranjan, Rajiv.
\newblock End-to-end {QoS} specification and monitoring in the internet of
  things.
\newblock {\em Newsletter, IEEE Technical Committee on Cybernetics for
  Cyber-Physical Systems}, 1(2), 2016.

\bibitem{chapterbook31}
P~Skene, James.
\newblock Language support for service-level agreements for application-service
  provision.
\newblock {\em Skene, J. (2007) Language support for service-level agreements
  for application-service provision. Doctoral thesis, University of London},
  2007.

\bibitem{chapterbook32}
Kyriakos Kritikos, Barbara Pernici, Pierluigi Plebani, Cinzia Cappiello, Marco
  Comuzzi, Salima Benrernou, Ivona Brandic, Attila Kertész, Michael Parkin,
  and P~Carro, Manue.
\newblock A survey on service quality description.
\newblock {\em ACM Computing Surveys (CSUR)}, 46(1):1, 2013.

\bibitem{sys206}
Fatima Alkandari and Richard~F. Paige.
\newblock Modelling and comparing cloud computing service level agreements.
\newblock In {\em Proceedings of the 1st International Workshop on Model-Driven
  Engineering for High Performance and CLoud Computing}, MDHPCL '12, pages
  3:1--3:6, New York, NY, USA, 2012. ACM.

\bibitem{chapterbook34}
Prem~Prakash Jayaraman, Karan Mitra, Saguna Saguna, Tejal Shah, Dimitrios
  Georgakopoulos, and P~Ranjan, Rajiv.
\newblock Orchestrating quality of service in the cloud of things ecosystem.
\newblock {\em In 2015 IEEE International Symposium on Nanoelectronic and
  Information Systems}, pages 185--190, 2015.

\bibitem{chapterbook35}
Xue Liu, Qixin Wang, Lui Sha, and P~He, Wenbo.
\newblock Optimal {QoS} sampling frequency assignment for real-time wireless
  sensor networks.
\newblock {\em in In RTSS}, 2003.

\bibitem{ch219}
Linlin Wu and Rajkumar Buyya.
\newblock Service level agreement ({SLA}) in utility computing systems.
\newblock In {\em Performance and dependability in service computing: Concepts,
  techniques and research directions}, pages 1--25. IGI Global, 2012.

\bibitem{awa}
A.~{Alqahtani}, Y.~{Li}, P.~{Patel}, E.~{Solaiman}, and R.~{Ranjan}.
\newblock End-to-end service level agreement specification for {IoT}
  applications.
\newblock In {\em 2018 International Conference on High Performance Computing
  Simulation (HPCS)}, pages 926--935 \color{black}, July 2018.

\bibitem{8}
Adriano Galati, Karim Djemame, Martyn Fletcher, Mark Jessop, Michael Weeks, and
  John McAvoy.
\newblock A ws-agreement based {SLA} implementation for the cmac platform.
\newblock In J{\"o}rn Altmann, Kurt Vanmechelen, and Omer~F. Rana, editors,
  {\em Economics of Grids, Clouds, Systems, and Services}, pages 159--171,
  Cham, 2014. Springer International Publishing.

\bibitem{10Buyya}
Redowan Mahmud and Rajkumar Buyya.
\newblock Fog computing: {A} taxonomy, survey and future directions.
\newblock {\em CoRR}, abs/1611.05539, 2016.

\bibitem{37}
Nicholas~J Dingle, William~J Knottenbelt, and Lei Wang.
\newblock Service level agreement specification, compliance prediction and
  monitoring with performance trees.
\newblock In {\em 22nd Annual European Simulation and Modelling Conference (ESM
  2008)}, pages 137--14, 2008.

\bibitem{chapterbook10}
Philip Bianco, Grace Lewis, and P~Merson, Paulo.
\newblock Service level agreements in service-oriented architecture
  environments.
\newblock {\em Technical Report CMU/SEI-2008-TN-021, Carnegie Mellon}, 2008.

\bibitem{zigbee}
A.~{Dementyev}, S.~{Hodges}, S.~{Taylor}, and J.~{Smith}.
\newblock Power consumption analysis of bluetooth low energy, zigbee and ant
  sensor nodes in a cyclic sleep scenario.
\newblock In {\em 2013 IEEE International Wireless Symposium (IWS)}, pages
  1--4, April 2013.

\bibitem{comparative}
Sakina Elhadi, Abdelaziz Marzak, Nawal Sael, and Soukaina Merzouk.
\newblock Comparative study of {IoT} protocols.
\newblock {\em Smart Application and Data Analysis for Smart Cities
  (SADASC'18)}, 2018.

\bibitem{cite6}
Pedro Garcia~Lopez, Alberto Montresor, Dick Epema, Anwitaman Datta, Teruo
  Higashino, Adriana Iamnitchi, Marinho Barcellos, Pascal Felber, and Etienne
  Riviere.
\newblock Edge-centric computing: Vision and challenges.
\newblock {\em SIGCOMM Comput. Commun. Rev.}, 45(5):37--42, September 2015.

\bibitem{5lit}
Rajkumar Buyya and Amir~Vahid Dastjerdi.
\newblock {\em Internet of Things: Principles and paradigms}.
\newblock Elsevier, 2016.

\bibitem{chapterbook28}
A~Rahmani, N~Thanigaivelan, T~Gia, J~Granados, B~Negash, P~Liljeberg, and
  P~Tenhunen, H.
\newblock Smart e-health gateway: Bringing intelligence to internet-of-things
  based ubiquitous healthcare systems.
\newblock {\em in 2015 12th Annual IEEE Consumer Communications and Networking
  Conference (CCNC)}, 2015.

\bibitem{ch3rPublishRate}
Guruduth Banavar, Tushar Chandra, Bodhi Mukherjee, Jay Nagarajarao, Robert~E
  Strom, and Daniel~C Sturman.
\newblock An efficient multicast protocol for content-based publish-subscribe
  systems.
\newblock In {\em Proceedings. 19th IEEE International Conference on
  Distributed Computing Systems (Cat. No. 99CB37003)}, pages 262--272. IEEE,
  1999.

\bibitem{ch3rBuffer}
Fatemeh Banaie, Jelena Misic, Vojislav~B Misic, Mohammad~Hossein Yaghmaee, and
  Seyed~Amin Hosseini.
\newblock Performance analysis of multithreaded {IoT} gateway.
\newblock {\em IEEE Internet of Things Journal}, 2018.

\bibitem{ch3Cloudonline}
Marcin Zukowski.
\newblock Cloud-based sql solutions for big data.
\newblock {\em Encyclopedia of Big Data Technologies}, pages 1--7, 2018.

\bibitem{ch3r4}
Deepak Vasisht, Zerina Kapetanovic, Jongho Won, Xinxin Jin, Ranveer Chandra,
  Sudipta Sinha, Ashish Kapoor, Madhusudhan Sudarshan, and Sean Stratman.
\newblock Farmbeats: An {IoT} platform for data-driven agriculture.
\newblock In {\em 14th $\{$USENIX$\}$ Symposium on Networked Systems Design and
  Implementation ($\{$NSDI$\}$ 17)}, pages 515--529, 2017.

\bibitem{ch3r8}
Jose Quevedo, Daniel Corujo, and Rui Aguiar.
\newblock Consumer driven information freshness approach for content centric
  networking.
\newblock In {\em 2014 IEEE conference on computer communications workshops
  (INFOCOM WKSHPS)}, pages 482--487. IEEE, 2014.

\bibitem{ch3r7}
Mohamed Ahmed~M Hail, Marica Amadeo, Antonella Molinaro, and Stefan Fischer.
\newblock On the performance of caching and forwarding in information-centric
  networking for the {IoT}.
\newblock In {\em International Conference on Wired/Wireless Internet
  Communication}, pages 313--326. Springer, 2015.

\bibitem{ch3r1}
Sandeep Pirbhulal, Heye Zhang, Md~E~Alahi, Hemant Ghayvat, Subhas Mukhopadhyay,
  Yuan-Ting Zhang, and Wanqing Wu.
\newblock A novel secure {IoT}-based smart home automation system using a
  wireless sensor network.
\newblock {\em Sensors}, 17(1):69, 2017.

\bibitem{ch3r3}
Prosanta Gope and Tzonelih Hwang.
\newblock Bsn-care: A secure {IoT}-based modern healthcare system using body
  sensor network.
\newblock {\em IEEE sensors journal}, 16(5):1368--1376, 2015.

\bibitem{ch3r1part2}
Ricardo Martinho and Dulce Domingos.
\newblock Quality of information and access cost of {IoT} resources in bpmn
  processes.
\newblock {\em Procedia Technology}, 16:737--744, 2014.

\bibitem{ch3r10}
S.~{Kolozali}, M.~{Bermudez-Edo}, D.~{Puschmann}, F.~{Ganz}, and P.~{Barnaghi}.
\newblock A knowledge-based approach for real-time {IoT} data stream annotation
  and processing.
\newblock In {\em 2014 IEEE International Conference on Internet of Things
  (iThings), and IEEE Green Computing and Communications (GreenCom) and IEEE
  Cyber, Physical and Social Computing (CPSCom)}, pages 215--222, Sep. 2014.

\bibitem{reffnote12}
Jean-Paul Calbimonte, Mehdi Riahi, Nikos Kefalakis, John Soldatos, and Arkady
  Zaslavsky.
\newblock Utility metrics specifications. openiot deliverable d422.
\newblock Technical report, {Infoscience, the École Polytechnique Fédérale
  de Lausanne (EPFL)}, 2014.

\bibitem{reff16}
Baoan Li and Jianjun Yu.
\newblock Research and application on the smart home based on component
  technologies and internet of things.
\newblock {\em Procedia Engineering}, 15(Supplement C):2087 -- 2092, 2011.
\newblock CEIS 2011.

\bibitem{reff22}
Cloud Standards~Customer Council.
\newblock Practical guide to cloud service agreements version 2.0.
\newblock Technical Report Supplement C, The Object Management Group (OMG),
  2011.

\bibitem{reff24}
E.~C. Kim, J.~G. Song, and C.~S. Hong.
\newblock An integrated cnm architecture for multi-layer networks with simple
  {SLA} monitoring and reporting mechanism.
\newblock In {\em Network Operations and Management Symposium, 2000. NOMS 2000.
  2000 IEEE/IFIP}, pages 993--994, 2000.

\bibitem{reff30}
Bhaskar Bhuyan, Hiren Kumar~Deva Sarma, Nityananda Sarma, Avijit Kar, and Rajib
  Mall.
\newblock Quality of service ({QoS}) provisions in wireless sensor networks and
  related challenges.
\newblock {\em Wireless Sensor Network}, 2(11):861, 2010.

\bibitem{reff36}
R.~Duan, X.~Chen, and T.~Xing.
\newblock A {QoS} architecture for {IoT}.
\newblock In {\em 2011 International Conference on Internet of Things and 4th
  International Conference on Cyber, Physical and Social Computing}, pages
  717--720, Oct 2011.

\bibitem{5}
R.~Ranjan.
\newblock Streaming big data processing in datacenter clouds.
\newblock {\em IEEE Cloud Computing}, 1(1):78--83, May 2014.

\bibitem{ch3rIngestion}
Luis Eduardo~Bautista Villalpando, Alain April, and Alain Abran.
\newblock Cloudmeasure: A platform for performance analysis of cloud computing
  systems.
\newblock In {\em 2016 IEEE 9th International Conference on Cloud Computing
  (CLOUD)}, pages 975--979. IEEE, 2016.

\bibitem{ch3rIngestionKa}
Tony Mancill.
\newblock Best practices for apache kafka.
\newblock \url{https://blog.newrelic.com/engineering/kafka-best-practices/},
  2018.
\newblock (Accessed on 07/03/2019).

\bibitem{ApacheKa}
Apache Kafka.
\newblock Apache kafka.
\newblock \url{https://kafka.apache.org/documentation/#gettingStarted}, 2017.
\newblock (Accessed on 07/03/2019).

\bibitem{AmazonK}
Amazon Web~Services AWS.
\newblock Amazon kinesis data firehose faqs.
\newblock \url{https://aws.amazon.com/kinesis/data-firehose/faqs/}, 2019.
\newblock (Accessed on 07/03/2019).

\bibitem{RabbitMQ}
Jack Vanlightly.
\newblock Rabbitmq vs kafka part 4 - message delivery semantics and guarantees
  — jack vanlightly.
\newblock
  \url{https://jack-vanlightly.com/blog/2017/12/15/rabbitmq-vs-kafka-part-4-message-delivery-semantics-and-guarantees},
  2017.
\newblock (Accessed on 07/04/2019).

\bibitem{19}
Shyamal Patel, Hyung Park, Paolo Bonato, Leighton Chan, and Mary Rodgers.
\newblock A review of wearable sensors and systems with application in
  rehabilitation.
\newblock {\em Journal of NeuroEngineering and Rehabilitation}, 9(1):21, 2012.

\bibitem{ch3rStream}
{Yuan Wei}, S.~H. {Son}, and J.~A. {Stankovic}.
\newblock Rtstream: real-time query processing for data streams.
\newblock In {\em Ninth IEEE International Symposium on Object and
  Component-Oriented Real-Time Distributed Computing (ISORC'06)}, pages 10
  pp.--, April 2006.

\bibitem{ch3rRajivStream}
Alireza Khoshkbarforoushha, Ritesh Ranjan, Raj Gaire, Prem~Prakash Jayaraman,
  John~G. Hosking, and Ehsan Abbasnejad.
\newblock Resource usage estimation of data stream processing workloads in
  datacenter clouds.
\newblock {\em ArXiv}, abs/1501.07020, 2015.

\bibitem{cite4}
Manuel D{\'\i}az, Cristian Mart{\'\i}n, and Bartolom{\'e} Rubio.
\newblock State-of-the-art, challenges, and open issues in the integration of
  internet of things and cloud computing.
\newblock {\em Journal of Network and Computer Applications}, 67:99--117, 2016.

\bibitem{17}
Nathan Marz and James Warren.
\newblock {\em Big Data: Principles and Best Practices of Scalable Realtime
  Data Systems}.
\newblock Manning Publications Co., Greenwich, CT, USA, 1st edition, 2015.

\bibitem{ch3Optimal}
Htway~Htway Hlaing, Hidehiro Kanemitsu, Tatsuo Nakajima, and Hidenori Nakazato.
\newblock On the optimal number of computational resources in mapreduce.
\newblock In {\em International Conference on Cloud Computing}, pages 240--252.
  Springer, 2019.

\bibitem{ch3verma2016big}
Ankush Verma, Ashik~Hussain Mansuri, and Neelesh Jain.
\newblock Big data management processing with hadoop mapreduce and spark
  technology: A comparison.
\newblock In {\em 2016 Symposium on Colossal Data Analysis and Networking
  (CDAN)}, pages 1--4. IEEE, 2016.

\bibitem{ch3Beyond}
Saeed Shahrivari.
\newblock Beyond batch processing: Towards real-time and streaming big data.
\newblock {\em Computers}, 3(4):117--129, 2014.

\bibitem{ch3Throghput}
Ahmed Eldawy and Mohamed~F Mokbel.
\newblock Spatialhadoop: A mapreduce framework for spatial data.
\newblock In {\em 2015 IEEE 31st international conference on Data Engineering},
  pages 1352--1363. IEEE, 2015.

\bibitem{das2017survey}
Kajaree Das and Rabi~Narayan Behera.
\newblock A survey on machine learning: concept, algorithms and applications.
\newblock {\em International Journal of Innovative Research in Computer and
  Communication Engineering}, 5(2):1301--1309, 2017.

\bibitem{ch3ML}
Meisong Wang, Prem~Prakash Jayaraman, Ellis Solaiman, Lydia~Y Chen, Zheng Li,
  Song Jun, Dimitrios Georgakopoulos, and Rajiv Ranjan.
\newblock A multi-layered performance analysis for cloud-based topic detection
  and tracking in big data applications.
\newblock {\em Future Generation Computer Systems}, 87:580--590, 2018.

\end{thebibliography}
\end{document}